\documentclass[a4paper, amsfonts, amssymb, amsmath, reprint, showkeys, twoside,notitlepage]{revtex4-2}
\def\apj{{ApJ}}                 % Astrophysical Journal
\def\mnras{{MNRAS}}             % Monthly Notices of the RAS
\def\prd{{Phys.~Rev.~D}}        % Physical Review D
\def\prl{{Phys.~Rev.~Lett.}}    % Physical Review Letters
\usepackage{amsmath,amstext}
\usepackage[T1]{fontenc}
\usepackage{amssymb}
\usepackage{graphicx}
\usepackage{ae,aecompl}
\usepackage{combelow}
\usepackage{graphicx}
\usepackage{tikz}
\usepackage{standalone}
\usepackage{hyperref}
\usepackage{amsmath}
\usepackage{amssymb}
\usepackage{mathtools}
\usepackage{bm}
\usepackage{cleveref}
\usepackage{tensor}
\usepackage{braket}
\usepackage{enumitem}
\usepackage{mhchem}
\usepackage{cancel}
\usepackage{tikz}
\usepackage{pgfplots}
\usetikzlibrary{external}
\usepackage{color}

\newcommand{\Kn}{\text{Kn}}

\def\be{\begin{equation}}
\def\ee{\end{equation}}
\def\beq{\begin{eqnarray}}
\def\eeq{\end{eqnarray}}

\newtheorem{definition}{Definition}

\begin{document}
\title{Infinite Order Hydrodynamics: an Analytical Example}
\author{ L.~Gavassino}
\affiliation{Department of Mathematics, Vanderbilt University, Nashville, TN, USA}

\begin{abstract}
We construct a kinetic model for matter-radiation interactions whose hydrodynamic gradient expansion can be computed analytically up to infinite order in derivatives, in the fully non-linear regime, and for arbitrary flows. The frequency dependence of the opacity of matter is chosen to mimic the relaxation time of a self-interacting scalar field. In this way, the transient sector simulates that of a realistic quantum field theory. The gradient series is found to diverge for most flows, in agreement with previous findings. We identify, for the model at hand, the mechanism at the origin of the divergence, and we provide a successful regularization scheme. Additionally, we propose a universal qualitative framework for predicting the breakdown of the gradient expansion of an arbitrary microscopic system undergoing a given flow. This framework correctly recovers all previously known instances of gradient expansion divergence. As a new prediction, we show that the gradient expansion diverges when the energy-dependent mean free path is unbounded above.
\end{abstract}

\maketitle

\textit{Introduction -} 
%\section*{Introduction}
The most pressing open question in relativistic fluid mechanics is: ``How far can we push hydrodynamics before it breaks down?'' \cite{Noronha-Hostler:2015wft,Strickland:2014pga,HellerAttractors2015,Alqahtani:2017mhy,Romatschke:2017acs,Romatschke2017,Romatschke2018FarFrom,FlorkowskiReview2018,DenicolNoronhaAttractor2019lio,HellerGradexp2021,Most:2021uck,MostAlfordNoronhaBulk2021zvc,HellerSingulant2022,Plumberg2022,MullinsDenicolCosmology,WagnerGavassino2023jgq,GavassinoFarFromBulk2023,GavassinoLargeFlux2023jyp,Aniceto:2024pyc,NoronhaSmallSystems2024dtq}

Let us make this question precise. The Knudsen number $\Kn=\lambda/L$ is the ratio between the particles' mean free path $\lambda$ (defining the microscopic scale) and the characteristic length scale $L$ of the flow of interest \cite{Hadjiconstantinou2006,Denicol2012Boltzmann}. It is common knowledge \cite{huang_book} that, if $\Kn \,{\rightarrow}\, 0$, we can model the system as an ideal fluid, while when $\Kn \gtrsim 1$, we need to rely on microphysics. Thus, there is some value of $\Kn \in (0,1]$ at which hydrodynamics stops working. Can we identify it precisely?

The answer depends on which hydrodynamic theory one is using. The ideal fluid breaks down at any finite value of $\Kn$, since it is non-dissipative, and it predicts that waves survive forever. These issues are fixed in Navier-Stokes theory, whose breakdown is conventionally set at $\Kn=0.1$ \cite{Struchtrup2011}. However, one may try to do even better. Most derivations of hydrodynamics from kinetic theory \cite{RochaBDNK2022ind,RochaReview2023ilf} and holography \cite{Baier2008,GrozdanovStuff2019}  express the stress-energy tensor $T^{\mu \nu}$ of a fluid as a formal power series in $\Kn$, known as the ``gradient expansion'' \cite{Struchtrup2011,Kovtun2019},
\begin{equation}\label{gradiunz}
    T^{\mu \nu}(\Kn)= T^{\mu \nu}_{\text{ID}}+T^{\mu \nu}_1 \Kn +T^{\mu \nu}_2 \Kn^2+...\, ,
\end{equation}
where the zeroth order is the ideal fluid, the first order is Navier-Stokes, the second order is the Burnett equation \cite{Struchup2005}, and higher orders correspond to fluid theories with higher derivative corrections. One hopes that, by adding more and more powers of $\Kn$, the regime of applicability of hydrodynamics will expand more and more, increasing the accuracy of hydrodynamics for all $\Kn$ up to the radius of convergence $\Kn^\star$ of the series \eqref{gradiunz}. Following this line of thought, it would then be natural to conclude that such radius $\Kn^\star$ is what ultimately marks the rigorous breakdown of hydrodynamics \cite{HellerHydroHedron2023jtd}.

Unfortunately, reality turns out to be more complicated. First, there are indications that, in most realistic scenarios, $\Kn^\star=0$ \cite{HellerGradexp2021,HellerSingulant2022,HellerBeyondBjorken2022}.  This means that, if we keep adding higher and higher orders in $\Kn$, at some point the region of applicability of \eqref{gradiunz} shrinks instead of expanding. Secondly, equation \eqref{gradiunz} makes sense only if the function $T^{\mu \nu}(\Kn)$ is analytic in $\Kn$. In principle, there may be non-smooth corrections like $\Kn^{3/2}$ \cite{KovtunStickiness2011}, or non-perturbative corrections like $e^{-1/\Kn}$ \cite{DenicolNoronhaDivergence2016bjh}. Therefore, the breakdown scale of hydrodynamics remains unknown.

The dream of any fluid theorist would be to have a microscopic model where $T^{\mu \nu}(\Kn)$ can be computed exactly for arbitrary flow, at arbitrary $\Kn$, in the fully non-linear regime, and where all the terms $T^{\mu \nu}_n$ in the series \eqref{gradiunz} have exact analytical formulas. Such model should have realistic interactions, to ensure that the dependence of $T^{\mu \nu}$ on $\Kn$ mimics the behavior of some microscopic quantum field theory. Finally, one should be able to extract from it general lessons about the expansion \eqref{gradiunz}. In this Letter, we provide a model that fulfills all these requirements.

Notation: We work in Minkowski space, with signature $(-,+,+,+)$, and adopt natural units: $c=\hbar=k_B=1$. 

\textit{The kinetic model -} 
%\section*{The kinetic model}
Our model setup is a radiation-hydrodynamic system \cite{Pomraning1973,mihalas_book,Farris2008,ShibataRadiation2011k,Sadowski2013,GavassinoRadiazione,RadiceBernuzziRadiation2023zlw}, namely a fluid mixture comprised of two substances: a material medium $M$ with negligibly short mean free path (i.e. $\Kn_M \equiv 0$), and a diluted radiation gas $R$ with a finite, possibly large, mean free path (i.e. $\Kn_R >0$) \cite{Thomas1930,Weinberg1971}. The former is modeled as an ideal fluid, with a well-defined temperature $T(x^\alpha)$ and flow velocity $u^\mu(x^\alpha)$. The latter can exist in far-from-equilibrium states, and we must track its kinetic distribution function $f_p(x^\alpha)$ \cite{UdeyIsrael1982,Thorne1981} ($p=\,$momentum). The total stress-energy tensor is the sum of a material non-viscous part and a radiation part:
\begin{equation}\label{fluxxuxx}
       T^{\mu \nu}= T^{\mu \nu}_M+\int \dfrac{d^3 p }{(2\pi)^3p^0} \, f_p \, p^\mu p^\nu \, . 
\end{equation}
The radiation particles do not interact with each other, but are randomly absorbed and emitted by the medium (neglecting scattering). Thus, Boltzmann's equation is of relaxation type \cite{Pomraning1973,AndersonWitting1974}. Taking the relaxation time to be $\tau_p{=}{-}u_\mu p^\mu/g$ (with $g{>}\,0\,{=}\,$const, for simplicity), we have
\begin{equation}\label{RRTTAA}
   p^\mu \partial_\mu f_p=-u_\mu p^\mu \,  \dfrac{f^{\text{eq}}_p{-}f_p}{\tau_p} =g \, (f^{\text{eq}}_p{-}f_p) \, .
\end{equation}
Since $\tau_p$ grows linearly with $-u_\mu p^\mu$, our medium is opaque at low frequencies, and transparent at high frequencies \cite{Pomraning1973}. The function $f^{\text{eq}}_p$ is the value that $f_p$ would have if the radiation was in local equilibrium, with the temperature $T$ and flow velocity $u^\mu$ of the medium. Depending on the statistics, one has $f^{\text{eq}}_p=(e^{-u_\mu p^\mu/T}{-}b)^{-1}$, $b \in \{0,\pm 1\}$ \cite{rezzolla_book}.
The backreaction of the radiation on the motion of the medium follows from the conservation law $\partial_\mu T^{\mu \nu}=0$ which, combined with \eqref{fluxxuxx} and \eqref{RRTTAA}, gives
\begin{equation}
    \partial_\mu T^{\mu \nu}_M = -g \int \dfrac{d^3 p }{(2\pi)^3p^0} \, (f^{\text{eq}}_p-f_p) \, p^\nu \, .
\end{equation}

\textit{Why this model? -} 
%\subsection*{Why this model?}
The relaxation time $\tau_p \propto \text{``energy''}$ mimics the transient behavior of a scalar field theory with $\phi^4$ interaction \cite{DenicolNoronhaRTA2022bsq}. This choice ensures that the gradient expansion \eqref{gradiunz} will exhibit ``realistic'' (i.e. QFT-inspired) scaling with powers of $\Kn_R=\text{max}\{\Kn_M,\Kn_R \}=\Kn$. Note that, in QCD plasmas, $\tau_p$ scales as a fractional power of the energy \cite{Dusling:2009df}, while the present model is closer to the kinetics of Yukawa's meson theory \cite{Srednicki:2007qs,GavassinoGapless2024rck}.

Our main reason for relying on radiation hydrodynamics (rather than dealing with ideal gases) is that it allows the introduction of momentum-dependent relaxation times, as in \eqref{RRTTAA}, without violating any conservation law \cite{RochaNovelRelaxation2021zcw}. Furthermore, within radiation hydrodynamics, the relaxation-time approximation is \textit{exact} (by Kirchhoff's law \cite{ClaytonBook1983}), if the medium has negligibly short equilibration timescale \cite{Pomraning1973,mihalas_book}. By contrast, in ideal gases, equations like \eqref{RRTTAA} are a pretty crude approximation of the Boltzmann collision integral.

Note that the fields $T$ and $u^\mu$ have unambiguous meanings in our model, as they characterize the local state of the medium, establishing a natural ``hydrodynamic frame'' \cite{Kovtun2019}. Indeed, providing initial data for the ``matter+radiation'' system requires specifying the initial states for both the radiation gas and the medium separately, so that $\{f_p,T,u^\mu\}$ are \textit{independent} dynamical degrees of freedom. This is convenient, since we don't need to worry about matching conditions \cite{RochaBDNK2022ind} (i.e., no integral constraint relates $T$ and $u^\mu$ to $f_p$ at a given time).

\textit{Linear response properties -} 
%\subsection*{Linear response properties}
Linearizing equations \eqref{fluxxuxx} and \eqref{RRTTAA}, and looking for solutions $\propto e^{ik_\mu x^\mu}$, we obtain 
\begin{equation}
\delta T^{\mu \nu}{=}\delta T^{\mu \nu}_M{+} \! \! \int \dfrac{d^3 p }{(2\pi)^3p^0} \, \dfrac{(f_{p,T}^{\text{eq}}\delta T{+}f_{p,u^\rho}^{\text{eq}}\delta u^\rho) \, p^\mu p^\nu}{1+ig^{-1}p^\alpha k_\alpha} \, .
\end{equation}
For $k^\alpha=(\omega,0,0,0)$, the integrand is singular whenever $\omega{=} - ig/p^0$, which approaches $\omega=0$ for $p^0 {\rightarrow} \infty$. Thus, the linear-response Green function of the stress-energy tensor has a branch cut that touches the origin of the complex $\omega-$plane. This is a known feature of $\phi^4$ \cite{MooreCuts2018mma,Rocha:2024cge} and other relativistic systems \cite{GavassinoGapless2024rck}, including, probably, QCD plasmas. Hence, the spectral properties of the present model agree with current expectations.

\textit{Analytic solution -} 
%\subsection*{Analytic solution}
Equations like \eqref{RRTTAA} can be solved analytically in the non-linear regime \cite{Pomraning1973}. Fixed an event $x^\alpha$ and a momentum $p^\alpha$, consider the worldline
\begin{equation}\label{geodesics}
    x^\alpha(\xi)=x^\alpha - \dfrac{p^\alpha}{g} \xi \, , 
\end{equation}
where $\xi$ is the optical depth. Travelling along \eqref{geodesics}, at fixed momentum $p^\alpha$, equation \eqref{RRTTAA} becomes an ordinary differential equation, $-\dot{f}_p(\xi)+f_p(\xi)=f_p^{\text{eq}}(\xi)$, which has unique solution for initial data at past infinity:
\begin{equation}\label{grizzo}
    f_p(x^\alpha)= \int_0^{+\infty} e^{-\xi} f^{\text{eq}}_p \bigg(x^\alpha {-} \dfrac{p^\alpha}{g} \xi \bigg) d\xi \, .
\end{equation}
Plugging \eqref{grizzo} in \eqref{fluxxuxx}, we obtain an \textit{exact} formula for the total stress-energy tensor:
\begin{figure}
\begin{center}
\includegraphics[width=0.51\textwidth]{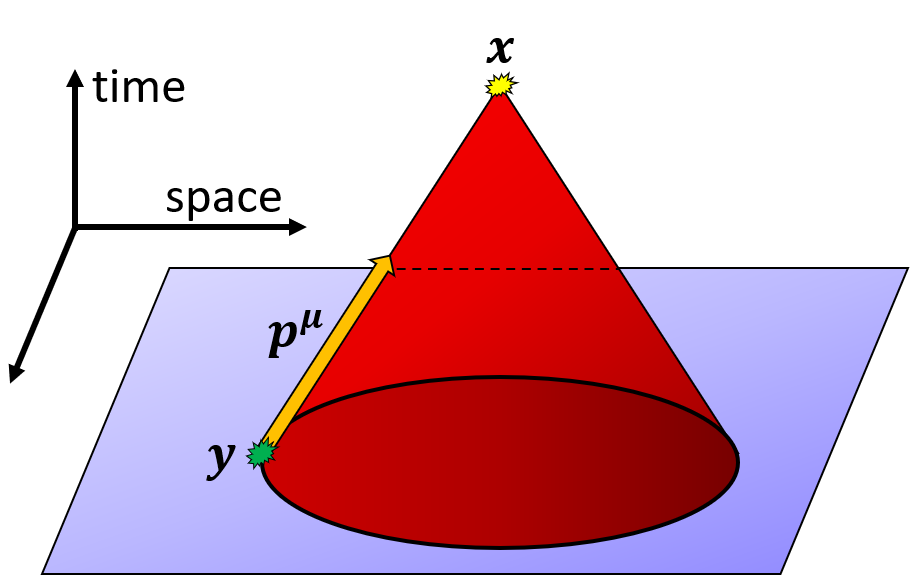}
	\caption{Visual representation of equation \eqref{fluxino}, viewed as a non-local constitutive relation $T^{\mu \nu}[T,u^\alpha]$. The value of $T^{\mu \nu}(x)$ receives contributions from \textit{all} events $y$ on the past lightcone of $x$ (although closer events contribute more). The information about $\{T(y),u^\alpha(y)\}$ is transported directly to $x$ by radiation particles emitted at $y$, which have probability $e^{-\xi}{=}e^{-g(x^0{-}y^0)/p^0}$ to reach $x$ before being absorbed.}
	\label{fig:constitutiverelx}
	\end{center}
\end{figure}
\begin{equation}\label{fluxino}
    T^{\mu \nu}= T^{\mu \nu}_M +\int \dfrac{d^3p}{(2\pi)^3 p^0} \,  p^\mu p^\nu \int_0^{+\infty} \! \! \! \!  f^{\text{eq}}_p \bigg(x^\alpha {-} \dfrac{p^\alpha}{g} \xi \bigg) \, e^{-\xi} \, d\xi \, .
\end{equation}
This is the function ``$T^{\mu \nu}(\Kn)$'' we were looking for. It remains valid at all orders in derivatives of the fluid fields $\{T,u^\mu\}$, which enter \eqref{fluxino} through $f^{\text{eq}}_p=(e^{-u_\mu p^\mu/T}-b)^{-1}$. We may interpret \eqref{fluxino} as an ``infinite-order hydrodynamic constitutive relation'', where $T^{\mu \nu}$ is a non-local functional of $T$ and $u^\mu$, see figure \ref{fig:constitutiverelx}. 

The ideal fluid is recovered in the infinite-opacity limit ($g\rightarrow +\infty$), since $x^\alpha(\xi)\rightarrow x^\alpha$, and the total stress-energy tensor $T^{\mu \nu}$ becomes that of a conglomerate fluid, where matter and radiation comove with velocity $u^\mu$ and both have temperature $T$:
\begin{equation}\label{iduzzuzuz}
\begin{split}
T^{\mu \nu}_{\text{ID}}={}& T^{\mu \nu}_M{+} \! \!\int \dfrac{d^3 p }{(2\pi)^3p^0} \, f^\text{eq}_p \, p^\mu p^\nu \\ ={}& (\varepsilon{+}P)u^\mu u^\nu{+}P\eta^{\mu \nu}\, .\\
\end{split}
\end{equation}
The energy density $\varepsilon(T,...)$ and the pressure $P(T,...)$ are the sum of matter and radiation parts in thermal equilibrium, where the matter contributions may depend on additional state variables besides $T$, including, possibly, some chemical potential.

\textit{Gradient series -} Let us evaluate the derivative expansion of \eqref{fluxino}. If $f^{\text{eq}}_p(x^\mu)$ is an entire function of $x^\mu$, we can express the integrand of \eqref{grizzo} as a Taylor series:
\begin{equation}\label{esperidi}
   f^{\text{eq}}_p (\xi)= \sum_{n=0}^{+\infty} \dfrac{\xi^n}{n!}\bigg(\dfrac{d}{d\xi}\bigg)^n f^{\text{{eq}}}_p(0) \, ,
\end{equation}
where $d/d\xi=-g^{-1}p^\alpha \partial_\alpha$ is the derivative along the parametric line \eqref{geodesics}. We note that $d/d\xi \sim \Kn$, due to the following chain of estimates ($L$ is the hydrodynamic scale):
\begin{equation}\label{knudso}
  \dfrac{d}{d\xi}=-\dfrac{p^\alpha \partial_\alpha}{g} \sim \dfrac{T}{gL} \sim \dfrac{-u^\mu p_\mu}{gL}=\dfrac{\tau_p}{L} \sim \dfrac{\lambda}{L}=\Kn .
\end{equation}
Hence, each $n^{\text{th}}$ term in the series \eqref{esperidi} corresponds to the power $\Kn^n$ in the Knudsen expansion.
Plugging \eqref{esperidi} into \eqref{fluxino}, and performing the integral in $\xi$, we can express the total stress-energy tensor as the sum of the ideal part \eqref{iduzzuzuz} and a viscous part, namely $T^{\mu\nu}=T^{\mu \nu}_{\text{ID}} +\Pi^{\mu \nu}$, with
\begin{equation}\label{MainMessage}
    \Pi^{\mu \nu} = \sum_{n=1}^{+\infty} (-g)^{-n} \partial_{\alpha_1}...\partial_{\alpha_n} \! \! \int \! \dfrac{d^3p}{(2\pi)^3  p^0} \, f^{\text{eq}}_p \, p^\mu p^\nu  p^{\alpha_1}...p^{\alpha_n} \, .
\end{equation}
This is the gradient series we were looking for. Again, no approximation was made, besides the assumptions of the kinetic model. The integrals in \eqref{MainMessage} are the moments of the equilibrium distribution function $f_p^{\text{eq}}$, all of which can be evaluated analytically. 

Let us discuss some notable properties of \eqref{MainMessage}:
\begin{itemize}
\item All the components of $\Pi^{\mu \nu}$ are in general non-zero, including $\Pi^{\mu \nu}u_\mu u_\nu$. Thus, the ``material frame'', where temperature and flow velocity are those of the medium, does not coincide with the Landau or the Eckart frame, as noted in \cite{Weinberg1971}.
%\item Space derivatives and time derivatives are treated on the same footing, and both enter with the same power-counting in $\Kn$.
\item All the contributions to \eqref{MainMessage} contain factors of the form $\partial_{(\mu_1} \partial_{\mu_2}...\partial_{\mu_k}\beta_{\nu)}$, with $\beta_\nu=u_\nu/T$. It follows that $\Pi^{\mu \nu}$ vanishes whenever $\beta_\nu$ is a Killing vector, namely $\partial_{(\mu}\beta_{\nu)}=0$, which is the condition of global thermodynamic equilibrium \cite{Israel_Stewart_1979,BecattiniBeta2016,GavassinoGibbs2021,GavassinoStabilityCarter2022}.
%\item The moments of $f^{\text{eq}}_p$ grow with $n$ like $(n{+}3)! T^{n+4}$. Thus, the coefficients $T^{\mu \nu}_n$ in the Knudsen series \eqref{gradiunz} are proportional to $(n{+}3)!$, making the gradient expansion \eqref{MainMessage} factorially divergent, as expected \cite{HellerSingulant2022}. 
\item If we truncate $\Pi^{\mu \nu}$ at finite order $n\geq 1$ (with Maxwell-Boltzmann statistics, i.e. $f^{\text{eq}}_p{=}e^{\beta_\rho p^\rho}$), the principal part of the conservation law $\partial_\mu T^{\mu \nu}=0$ (regarded as a dynamical equation for $\beta_\nu$) is 
\begin{equation}
 \dfrac{1}{ (-g)^{n}}\int \! \dfrac{f^{\text{eq}}_p \, d^3p}{(2\pi)^3  p^0} \,  \, p^\mu p^\nu p^\rho  p^{\alpha_1}...p^{\alpha_n} \partial_{\alpha_1}...\partial_{\alpha_n} \partial_\mu\beta_{\rho} \, .
\end{equation}
The associated characteristic polynomial \cite{CourantHilbert2_book} is
\begin{equation}
   \mathcal{P}(\zeta_\alpha)= \det\bigg[ \int \! \dfrac{ f^{\text{eq}}_p \, d^3p}{(2\pi)^3  p^0} \,  (p^\alpha \zeta_\alpha)^{n+1} p^\nu p^\rho \bigg] \, .
\end{equation}
For $n$ odd, the matrix in the determinant is positive definite for all $\zeta_\alpha$. Thus, the evolution equation is elliptic, making all odd-order truncations acausal, unstable, and ill-posed \cite{Rauch_book,GavassinoCasmir2022,GavassinoSuperluminal2021}.
\end{itemize}

\textit{Divergence of the gradient series -} Let us show that \eqref{MainMessage} is usually divergent. We can estimate the order of magnitude of each term of the series by making the formal replacements $\partial_\alpha \rightarrow 1/L$ and $f_{p}^{\text{eq}}\rightarrow e^{-p^0/T}$. Then, equation \eqref{MainMessage} becomes (ignoring constant factors)
\begin{equation}\label{divergendo}
    \dfrac{\Pi^{\mu \nu}}{T^4} \sim \sum_{n}  \bigg( \dfrac{T}{gL}\bigg)^n (n{+}3)! \, ,
\end{equation}
which is factorially divergent. 

Let us now repeat the above calculation, assuming that the medium cannot produce radiation particles whose energy exceeds some large cutoff scale $\Lambda \gg T$, thereby restricting integrals in $d^3 p$ to the sphere $p^0 \leq \Lambda$. Then, the series \eqref{divergendo} becomes
\begin{equation}
   \dfrac{\Pi^{\mu \nu}}{T^4} \sim \sum_n \bigg( \dfrac{T}{gL}\bigg)^n \gamma(n{+}4,\Lambda/T) \, ,
\end{equation}
where $\gamma$ is the lower incomplete Gamma function, which becomes $(n+3)!$ if we send $\Lambda\rightarrow +\infty$ at fixed $n$. Instead, let us evaluate the series for \textit{finite} $\Lambda \gg T$. At large $n$, the addends scale like
\begin{equation}
   \bigg( \dfrac{T}{gL}\bigg)^n \gamma(n{+}4,\Lambda/T) \sim  \bigg( \dfrac{\Lambda}{gL}\bigg)^n \dfrac{e^{-\Lambda/T}(\Lambda/T)^4}{n} \, . 
\end{equation}
If $gL>\Lambda$, the series converges. Furthermore, the large-$n$ part of the series adds up to
\begin{equation}\label{estimiamo}
   - e^{-\Lambda/T}\bigg( \dfrac{\Lambda}{T} \bigg)^4\ln\bigg(1-\dfrac{\Lambda}{gL}\bigg) \, ,
\end{equation}
which is negligible in the limit $\Lambda \gg T$.

The above argument shows that the divergence of the gradient expansion comes from fictitious particles that carry unreasonably large energy. We can prove that such particles are unphysical by showing that the exact formula \eqref{fluxino} is unaffected by the cutoff $\Lambda$. Called $T^{\mu \nu}_\Lambda$ the regularised stress-energy tensor, we have
\begin{equation}\label{fluxinoCutoff}
\begin{split}
    \dfrac{T^{\mu \nu} {-}T^{\mu \nu}_\Lambda}{T^4} & {=}  \int_{p^0 \geq \Lambda} \dfrac{d^3p}{(2\pi)^3 p^0} \,  \dfrac{p^\mu p^\nu }{T^4}\int_0^{+\infty} \! \! \! \!  f^{\text{eq}}_p (\xi) \, e^{-\xi} \, d\xi \\
    & {\sim} \int_{p^0 \geq \Lambda} \dfrac{d^3p}{(2\pi)^3} \,  \dfrac{p^0 e^{-p^0/T} }{T^4}{\sim} \bigg( \dfrac{\Lambda}{T} \bigg)^3 e^{-\Lambda/T} \, ,
\end{split}
\end{equation}
which tends to zero when $\Lambda {\gg} T$. This estimate shows that the cutoff $\Lambda$ can be used as a practical tool, to regularize the gradient expansion \eqref{MainMessage}, and restore agreement with \eqref{fluxino}. One only needs to make sure that $gL > \Lambda \gg T$. Unfortunately, this also requires $\Kn=T/(gL)\ll 1$. This suggests that adding higher derivative terms may not significantly expand the regime of applicability of hydrodynamics. Instead, it may just lead to a refinement of the accuracy of hydrodynamics, within the same interval $0 \leq \Kn \lesssim 0.1$ where Navier-Stokes is already valid. Indeed, several previous studies seem to support the present conclusion \cite{Geroch2001,Struchup2005,GarciaColin2008,WagnerGavassino2023jgq}.

\textit{A quick example -} We can check the consistency of the estimates \eqref{estimiamo} and \eqref{fluxinoCutoff} with an example. Take $u^\mu=(1,0,0,0)$ and $T(t)\propto 1-a\sin(t/L)$, with $a \rightarrow 0$. Linearising in $a$, and defining $\Kn=T/(gL)$, the value of $\Pi^{00}(0)$ according to \eqref{fluxino} and \eqref{MainMessage} is given by, respectively,
\begin{figure}
\begin{center}
\includegraphics[width=0.47\textwidth]{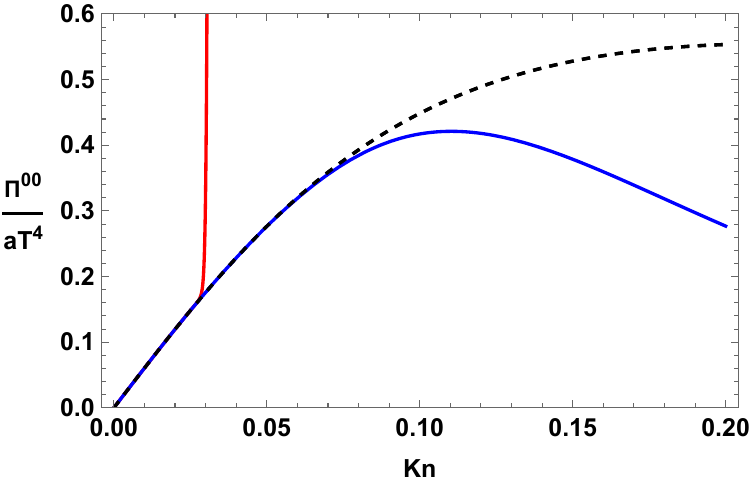}
\includegraphics[width=0.47\textwidth]{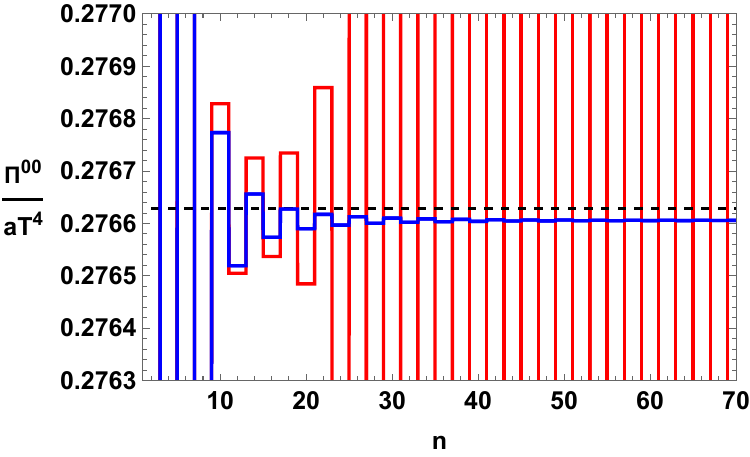}
	\caption{Upper panel: Exact viscous density $\Pi^{00}(t{=}0)$ according to \eqref{two} (dashed) compared to the non-regularised gradient series \eqref{three} (red) truncated at $n{=}70$, and the regularised series \eqref{four} (blue) with $n \rightarrow \infty$ and cutoff scale $\Lambda/T=(\Kn{+}0.01)^{-1}$. Lower panel: Convergence test at $\Kn{=}0.05$ (same colors).}
	\label{fig:cutoff}
	\end{center}
\end{figure}
\begin{flalign}
\text{Exact:} && \label{two} \dfrac{\Pi^{00}}{aT^4} &{=} \int_0^{+\infty} \dfrac{dz}{2\pi^2}  \dfrac{\Kn \, z^5 e^{-z}}{1+\Kn^2 \, z^2} \,,&\\
\text{Expand:} && \label{three} \dfrac{\Pi^{00}}{aT^4} &{=} \sum_{n \text{ odd}} \dfrac{(-1)^{\frac{n{-}1}{2}}}{2\pi^2} \Kn^n (n{+}4)!  \,,&\\
\text{Cutoff:} && \label{four} \dfrac{\Pi^{00}}{aT^4} &{=} \sum_{n \text{ odd}} \dfrac{(-1)^{\frac{n{-}1}{2}}}{2\pi^2} \Kn^n \gamma\bigg(n{+}5,\dfrac{\Lambda}{T}\bigg) \, ,
\end{flalign}
where equation \eqref{four} is the same as \eqref{three}, corrected with cutoff $\Lambda/T\,{<}\,\Kn^{-1}$ (for convergence of the series). We observe in figure \ref{fig:cutoff} the breakdown of the gradient expansion at order 70 (higher orders explode at smaller $\Kn$). The cutoff $\Lambda$ restores agreement with \eqref{fluxino} at all orders, within an error that survives at large $n$ and increases with $\Kn$. This fixes the breakdown of the regularised gradient series at $\Kn \sim 0.1$, which is precisely when Navier-Stokes ceases to hold \cite{Struchtrup2011}.

\textit{Origin of the divergence -} We now provide a general explanation for the divergence of the gradient expansion, which applies to the present model and to most other models in the literature.

Let $A$ be a density of interest, located at $x$. By causality, the event $x$ receives influence from all events $y$ in the causal past $J^-(x)$ of $x$ \cite{Wald}. Information is transported from $y$ to $x$  by messengers (e.g. particles or excitations), each having an associated energy (or frequency) $E$. Thermalization erases the messenger's memory about $y$ over a timescale $\tau(E)$. Thus, we can schematically write
\begin{equation}\label{sblurf}
    A = \int_0^{+\infty} \! \! \! dE \int_{J^-(x)} \! \! \! d^4 y \, \,  \sigma_A(E,y) e^{-(x^0-y^0)/\tau(E)} \, ,
\end{equation}
where $\sigma_A(E,y)$ quantifies how many messengers with energy $E$ are emitted from $y$, and ``how much $A$'' they carry. The Knudsen expansion \eqref{esperidi} corresponds, in this general setting, to a Taylor expansion of $\sigma_A(E,y)$ around $x$. Since, in the lightcone, $|x^j-y^j| \leq x^0-y^0$, we may write, schematically (recall: $\partial_\alpha \rightarrow 1/L$),
\begin{equation}\label{grbruf}
    \sigma_A(E,y)\sim \sum_{n} \dfrac{\sigma_A(E,x)}{n!} \bigg(\dfrac{x^0-y^0}{L}\bigg)^n \, .
\end{equation}
Plugging \eqref{grbruf} into \eqref{sblurf}, and taking the series outside the volume integral, we obtain, up to $n-$independent factors,
\begin{equation}\label{twntyfourrr}
    A \sim \int_0^{+\infty} \! \! \! dE \, \sigma_A(E,x) \sum_n \bigg[\dfrac{\tau(E)}{L} \bigg]^{n+1} \, .
\end{equation}
Now we see the problem: If $\tau(E)$ is unbounded above, fixed any $L$, there is some energy $E$ such that $\tau(E)>L$, and the geometric series in \eqref{twntyfourrr} diverges for such $E$. This implies that we were not allowed to exchange the series in $n$ with the integral in $d^4 y$, and in general \cite{Rudin_book}
\begin{equation}
    A\bigg[ \sum_{n=0}^{+\infty} \sigma_{A,n} \bigg] \neq \sum_{n=0}^{+\infty} A[\sigma_{A,n}] \, .
\end{equation}

This is precisely the issue with our analytical model. The relaxation time $\tau_p=-u_\mu p^\mu/g$ diverges at high energies, and we cannot take the series \eqref{esperidi} outside the integral \eqref{fluxino}. The introduction of the energy cutoff resolves the problem because, if $\Lambda/T<\Kn^{-1}$, the series \eqref{twntyfourrr} converges for all $E\leq \Lambda$, since (see figure \ref{fig:constitutiverelx})
\begin{equation}
    \dfrac{\tau(E)}{L}= \dfrac{E}{gL} \leq \dfrac{\Lambda}{gL}=\dfrac{\Lambda}{T} \, \Kn <1 \, .
\end{equation}
The same conclusions apply to all kinetic models of QCD plasmas \cite{Dusling:2009df,Dusling:2011fd,Kurkela:2017xis}, whose $\tau(E)$ ($\propto E^{q}$, with $q{>}0$) diverges at large $E$ (i.e. for jets). According to \eqref{twntyfourrr}, the gradient series of all such quark-matter models has a jet-induced divergence, since, fixed $L$, $\tau(E){\gg}L$ when $E {\rightarrow} \infty$.

With similar reasoning, we can also predict the breakdown of all gradient expansions due to stochastic fluctuations \cite{KovtunStickiness2011}. In fact, long-wavelength hydrodynamic fluctuations can transport information from $y$ to $x$, and have decay timescale $\tau(E)\sim E^{-2}$ (with $E{=}$frequency), which diverges at small $E$. This generates an infrared (i.e. low $E$) divergence in \eqref{twntyfourrr}, consistently with \cite{KovtunStickiness2011}.

\newpage

In the Supplementary Material \footnote{See Supplemental Material for some general considerations about the gradient expansion in linear-response theory. The Supplementary Material also
includes References \cite{Steffe_book,jackson_classical_1999,HellerBounds2023,GavassinoBounds2023,Denicol_Relaxation_2011,GavassinoBurgers2023}.}, the above claims are shown to hold (as rigorous statements) within linear-response theory.

\textit{General flows -} In the above estimates, we always assumed that $(\partial)^n \rightarrow 1/L^{n}$. This means that our results apply to sinusoidal and exponential flows, or finite superpositions thereof. However, different flows attribute different weights to each derivative. In Bjorken expansion \cite{BjorkenFlow1983}, where everything scales like $(y^0)^{-1}$, the correct formal replacement is $(\partial)^n \, {\rightarrow} \, n!/L^n$, and \eqref{twntyfourrr} becomes
\begin{equation}\label{twntyfourrr2}
    A \sim \int_0^{+\infty} \! \! \! dE \, \sigma_A(E,x) \sum_n \bigg[\dfrac{\tau(E)}{L} \bigg]^{n+1} n! \, ,
\end{equation}
which diverges independently from the value of $\tau(E)$. This explains why Israel-Stewart hydrodynamics has divergent gradient expansion in Bjorken flow \cite{Romatschke2018FarFrom}, despite having only one relaxation time. We also understand why Borel summation gives correct predictions in Bjorken flow \cite{HellerAttractors2015}. By Borel-summing the series \eqref{twntyfourrr2}, we obtain\footnote{The principal value guarantees that the (unphysical) singularity at $t=L$ gives a finite contribution to the integral. When $L \gg \tau(E)$, we can neglect it, and the integral can be effectively stopped at $L-\tau(E)$.}
\begin{equation}
    \sum_{n=0}^{+\infty} \bigg[\dfrac{\tau(E)}{L} \bigg]^{n+1} n! \, \, \stackrel{\mathcal{B}}{=} \, \, \text{P.V.}\! \! \int_0^{+\infty} \dfrac{e^{-t/\tau(E)}}{L-t}  dt \, ,
\end{equation}
which has precisely the form of the spacetime integral in \eqref{sblurf}, with $\sigma_A(y)$ that decays like $1/y^0$. This suggests that Borel summation ``undoes'' the gradient expansion, whose existence wasn't justified in the first place, and it reconstructs the original (correct) integral \eqref{sblurf} from which the gradient expansion was (improperly) derived.

The same reasoning can be generalized to arbitrary flows. For example, suppose that the Taylor expansion of $\sigma_A(y)$ has finite radius of convergence $\mathcal{R}$. Then, for large $n$, we have, schematically, $(\partial)^n \rightarrow n!/\mathcal{R}^n$ \cite{AnalyticFunctions_book}, and 
\begin{equation}\label{twntyfourrr3}
    A \sim \int_0^{+\infty} \! \! \! dE \, \sigma_A(E,x) \sum_n \bigg[\dfrac{\tau(E)}{\mathcal{R}} \bigg]^{n+1} n! \, .
\end{equation}
This explains why, in most realistic flows, where $T(x^\alpha)$ and $u^\mu(x^\alpha)$ are not entire functions, the gradient expansion is factorially divergent \cite{HellerBeyondBjorken2022}. The general validity of this argument within linear-response theory is demonstrated in the Supplementary Material.

%For a given flow, one can define a sequence $(\partial)^n \rightarrow a_n/L^n$. Then, depending on the behavior of $a_n$, one can ``predict'' (modulo exact cancellations) whether the given flow has a convergent gradient expansion. For example, Couette flow has $a_n\sim \{1,1,0,0,0,..\}$, resulting in a convergent expansion. On the other hand, an accretion disk revolving with Keplerian velocity $u(r) \sim 1/\sqrt{r}$ has $a_n \sim \Gamma(1/2{+}n)$, and the expansion diverges.

\textit{Conclusions -} Our analysis reveals what infinite-order hydrodynamics looks like. The stress-energy tensor is still a functional of the fluid fields $\{T,u^\alpha\}$, just like in finite-order hydrodynamics. However, such a functional $T^{\mu \nu}[T,u^\alpha]$ is \textit{non-local} at infinite order\footnote{The non-local nature of equations involving an infinite series of derivatives is well-known \cite{GavassinoDispersion2024}. Take, e.g., the constitutive relation $\varphi(x)=e^{a\partial_x}\psi(x)$, which is equivalent to $\varphi(x)=\psi(x+a)$ \cite{Callen_book}.}. By causality, $T^{\mu \nu}(x)$ depends solely on the restriction of $T$ and $u^\alpha$ within the past lightcone of $x$. Furthermore, by dissipation, it is mostly affected by past events that are close to $x$, as it loses memory of the far past. 

Starting from these observations, we proposed a simple explanation of why the gradient series \eqref{gradiunz} often diverges:
In a nutshell, the construction of the gradient series amounts to interchanging the Taylor series of the fields $\{T,u^\alpha\}$ with the spacetime integral defining $T^{\mu \nu}[T,u^\alpha]$. This delicate exchange of limits is often not allowed. In fact, the $n^\text{th}$ term of a Taylor expansion diverges like $\sim t^n$ at past infinity, and the usual theorems for exchanging limits and integrals (like the dominated convergence theorem \cite{Rudin_book}) do not apply. In some cases, Borel resummation ``brings the series back inside $T^{\mu \nu}$'', and it reconstructs the correct functional $T^{\mu \nu}[T,u^\alpha]$.

So, when does hydrodynamics break down? From our analytical example, one can give two answers, depending on the definition of the word ``hydrodynamics''.
\begin{definition}[Broad]\label{Broad}
Hydrodynamics is the ability to express the conserved currents in terms of the fluid variables alone, e.g. $T^{\mu \nu}=T^{\mu \nu}[T,u^\alpha]$.
\end{definition}
Using this definition, we find that hydrodynamics never breaks down, at any $\Kn$, provided we know the near past of the fluid (``near past'' $\approx$ ``few mean free paths ago''). Unfortunately, according to definition \ref{Broad}, solving hydrodynamics amounts to solving $\partial_\mu T^{\mu \nu}[T,u^\alpha]=0$, which is, in general, a convoluted integrodifferential equation. 
\begin{definition}[Narrow]\label{Narrow}
Hydrodynamics is the ability to express the conserved currents as local functions of the fluid fields and a finite number of derivatives thereof, e.g. $T^{\mu \nu}=T^{\mu \nu}(T,u^\alpha, \partial T, \partial u^\alpha,...,\partial^n T,\partial^n u^\alpha)$.
\end{definition}
In our model, nature conspires to break the validity of the whole gradient expansion at the same breakdown scale of Navier-Stokes ($\Kn \sim 0.1$). Thus, higher derivative corrections do not expand the applicability of the gradient expansion beyond first order. \\

\section*{Acknowledgements}
This work was partially supported by a Vanderbilt Seeding Success Grant. We also Thank M. Disconzi and J. Noronha for reading the manuscript and providing useful feedback.

%``

\appendix

\bibliography{Biblio}

\newpage

\onecolumngrid
\newpage
\begin{center}
  \textbf{\large Infinite Order Hydrodynamics: An Analytical Example\\Supplementary Material}\\[.2cm]
  L. Gavassino,$^{1}$\\[.1cm]
  {\itshape ${}^1$Department of Mathematics, Vanderbilt University, Nashville, TN, USA\\}
(Dated: \today)\\[1cm]
\end{center}

\setcounter{equation}{0}
\setcounter{figure}{0}
\setcounter{table}{0}
\setcounter{page}{1}
\renewcommand{\theequation}{S\arabic{equation}}
\renewcommand{\thefigure}{S\arabic{figure}}

Here, we formally derive some of the claims about the divergence of the gradient expansion that were demonstrated, on qualitative grounds, in the main text. To make the problem analytically tractable (and the derivation rigorous), we must restrict our analysis to the linear regime.

\subsection{Derivation of equation (25)}

The properties of the linearised constitutive relations of a fluid can be discussed using linear response theory \cite{Denicol_Relaxation_2011}. For example, the response of a stress to an applied strain rate is used in rheology to characterize the mechanical properties of a material medium \cite{Steffe_book}. Given the general applicability of this framework, we will use it as our playground for testing the claims of the main text.

Suppose that a fluid is perturbed by an external force $F(t)$, which triggers a response in an observable $\Pi(t)$. For clarity, we assume that both quantities $F$ and $\Pi$ depend only on one parameter $t= \, \,$``time'', although it should be possible to generalize the results below to a setting where an additional explicit dependence on $\textbf{x}$ is included\footnote{Indeed, the assumption that the quantities depend only on time does not necessarily imply that the perturbation is assumed homogenous. For example, one may assume that the force acts at a fixed wavenumber $\textbf{k}$, so that $F(t)\equiv F(t,\textbf{k})$ and $\Pi(t)\equiv \Pi(t,\textbf{k})$.}. Within linear response theory, the most general (time-translation-invariant) relation linking $\Pi$ to $F$ takes the form
\begin{equation}\label{andiamo!!!!}
    \Pi(t)= -\int_{-\infty}^{+\infty} G(t-t') F(t') dt' \, ,
\end{equation}
where $G$ is a linear-response Green function. Causality and stability requirements \cite{jackson_classical_1999,HellerBounds2023,GavassinoBounds2023} imply that the Fourier transform $G(\omega)$ is analytic in the upper complex $\omega$-plane, which is true if and only if $G(t)$ vanishes for $t < 0$. For $t \geq 0$, we can close the contour integral in the lower $\omega$-plane, and shrink the path around the singularities of $G(\omega)$. Thus, introducing a complex (Radon-Nikodym-type) measure $d\lambda$, which runs over all singularities of $G(\omega)$ and quantifies their magnitude, we can rewrite integral $G(t)=(2\pi)^{-1}\int d\omega G(\omega) e^{-i\omega t}$ in the following equivalent form:
\begin{equation}\label{convincitio}
    G(t)=\Theta(t) \int e^{-i\omega(\lambda) t}d\lambda \, ,
\end{equation}
where $\omega(\lambda)$ is the location of the singularity in the complex $\omega$-plane. Plugging \eqref{convincitio} into \eqref{andiamo!!!!}, and changing integration variable to $\xi=t-t'$, we obtain
\begin{equation}
\Pi(t)=-\int d\lambda \int_0^{+\infty} d\xi \, e^{-i\omega(\lambda)\xi}F(t-\xi) \, .    
\end{equation}
Expanding $F(t-\xi)$ in Taylor series around $\xi=0$, formally bringing the series outside the integral in $d\xi$, and integrating in $\xi$, we finally arrive at the desired equation:
\begin{equation}\label{labone}
\Pi(t) = \int d\lambda \sum_{n=0}^{+\infty} \dfrac{(-1)^{n+1} F^{(n)}(t)}{\big[ i\omega(\lambda)\big]^{n+1}}  \, .  
\end{equation}
To see that this equation is analogous to equation (25) of the main text, suppose that the external force has a sinusoidal form, namely $F(t)\propto e^{-i\omega_0 t}$, for some frequency $\omega_0\in \mathbb{R}$. Then, $F^{(n)}(t)=(-i\omega_0)^n F(t)$, and equation \eqref{labone} becomes
\begin{equation}\label{w0}
\Pi(t)=- \int d\lambda \dfrac{F(t)}{i\omega_0} \sum_{n=0}^{+\infty} \bigg[\dfrac{\omega_0}{\omega(\lambda)} \bigg]^{n+1} \, ,
\end{equation}
which has precisely the form of equation (25). Indeed, all the general facts about the gradient expansion that we discussed in the main text can be argued directly from equation \eqref{labone}.

\newpage

\subsection{Application to kinetic theory}

Linear-response theory is a standard technique for deriving linear constitutive relations in a dilute gas \cite{Denicol_Relaxation_2011,GavassinoBurgers2023,WagnerGavassino2023jgq,Rocha:2024cge}. Usually, one identifies $\Pi$ with a relevant stress component, and $F$ with some gradient of the flow. Let us see, then, where equation \eqref{w0} leads us. 

According to the analysis carried out in \cite{GavassinoGapless2024rck}, if the mean-free path $\tau(p^0)$, expressed as a function of the energy $p^0$, is unbounded above, the non-hydrodynamic sector is necessarily gapless. In other words, in the integral \eqref{convincitio}, there exists some region in $\lambda$-space (with non-zero $\lambda$-measure) for which $\mathfrak{Im} \, \omega(\lambda)\rightarrow 0$. Hence, by looking at equation \eqref{w0}, we see that, for any $\omega_0 \neq 0$, the series diverge on a set of $\lambda$ values with finite $\lambda$-measure.

In summary: If a system possesses particles with arbitrarily long mean free path, the expansion of $\Pi(t)$ in derivatives of $F(t)$ is necessarily divergent (at least absolutely\footnote{I.e., modulo fine-tuned cancellations within the $\lambda$-integral.}) along all Fourier modes, with the only exception of $\omega_0 \equiv 0$.

\subsection{Divergence of the gradient expansion for non-entire flows}

In the main text, we argued that the gradient expansion is likely to be factorially divergent when the fluid variables have a finite radius of convergence in spacetime. Here, we provide a demonstration of this claim with a simple example. 
Take
\begin{equation}\label{ft}
    F(t)= \dfrac{1}{1+t^2} \, .
\end{equation}
This function is perfectly smooth, being of class $C^{\infty}$. Furthermore, its Fourier transform is well defined, being proportional to $e^{-|\omega|}$. Nevertheless, if we expand \eqref{ft} around, say, $t=0$, its radius of convergence equals $1$, which is a finite number. Thus, the gradient expansion of $\Pi(0)$ should be factorially divergent, according to our discussion in the main text. Indeed, this is precisely what happens. Evaluating the linear-response relation \eqref{labone} explicitly, we obtain
\begin{equation}
\Pi(0) = \int d\lambda \sum_{n=0}^{+\infty} \dfrac{(-1)^{n+1}  (2n)!}{\big[i\omega(\lambda)\big]^{2n+1}} \, ,
\end{equation}
which has the same form as equation (30) of the main text, and diverges factorially.

\newpage

\label{lastpage}

\end{document}